\documentclass[11pt]{article}
\usepackage{ijcai18}
\usepackage{times}
\usepackage{helvet}
\usepackage{courier}
\usepackage{latexsym}
\usepackage{amsfonts,amssymb}
\usepackage{amsmath}
\usepackage{graphicx}
\usepackage{algorithm}
\usepackage{algorithmic}
\usepackage{float}
\usepackage{array}
\usepackage{multirow}
\usepackage{url}

\date{}

\title{SAR: Semantic Analysis for Recommendation}
\date{}

\author{Han Xiao$^\spadesuit$, Lian Meng \\
	State Key Lab. of Intelligent Technology and Systems, \\ 
	National Lab. for Information Science and Technology, \\
	Dept. of Computer Science and Technology, Tsinghua  University, Beijing 100084, PR China
	\\ Almighty.Xiao.Han@iCloud.com; Mengl15@Foxmail.com; 
	\\$^\spadesuit$Correspondence Author: \url{http://www.semantics.top}
}

\begin{document}
\maketitle

\begin{abstract}
Recommendation system is a common demand in daily life and matrix completion is a widely adopted technique for this task. However, most matrix completion methods lack semantic interpretation and usually result in weak-semantic recommendations. To this end, this paper proposes a \textbf{S}emantic \textbf{A}nalysis approach for \textbf{R}ecommendation systems \textbf{(SAR)}, which applies a two-level hierarchical generative process that assigns semantic features and categories for user/item. SAR learns the semantic representations of users/items merely from user ratings on items, which offers a novel path to recommendation by semantic matching with the learned semantic representations. Extensive experiments demonstrate SAR outperforms other state-of-the-art baselines substantially.
\end{abstract}

\section{Introduction}
As a necessary demand in daily life, recommendation systems could benefit various applications, such as information retrieval, question answering, sentiment analysis, and more. To mathematically formulate recommendation, rating matrix is focused by many researches, whose entry $M_{u,i}$ indicates the rating of item $i$ made by user $u$. However, there are many unknown entries in this matrix and it is the goal of rating-based recommendation to predict these missing values. Until now, many efforts have been attempted, such as low-rank based methods \cite{Lee2016LLORMA}, latent space models \cite{Song2015Bayesian}, collaborative filtering \cite{Rao2015Collaborative}, norm constrained methods \cite{Cai2013A}, deep learning \cite{Zhang2006Employing}, social integration \cite{Liu2016Learning}, etc.

Among these models, matrix completion methods (e.g. matrix factorization) is one of the most popular branches. Specifically, this branch models the rating matrix with various assumptions such as low-rank. Further, under the specific assumptions, an optimization problem is formulated to complete the rating matrix for recommendation. But, due to the lack of underlying semantics, the accuracy of this branch is unsatisfactory in real-world applications. To incorporate more external resources, social information such as user relationships can be integrated into matrix completion \cite{Liu2016Learning}. However, social information is also noisy and only
offers limited performance improvement, which is still not a panacea.

\begin{figure*}[t]
	\centering
	\label{CoreIdeaFigure}
	\includegraphics[width=0.80\linewidth]{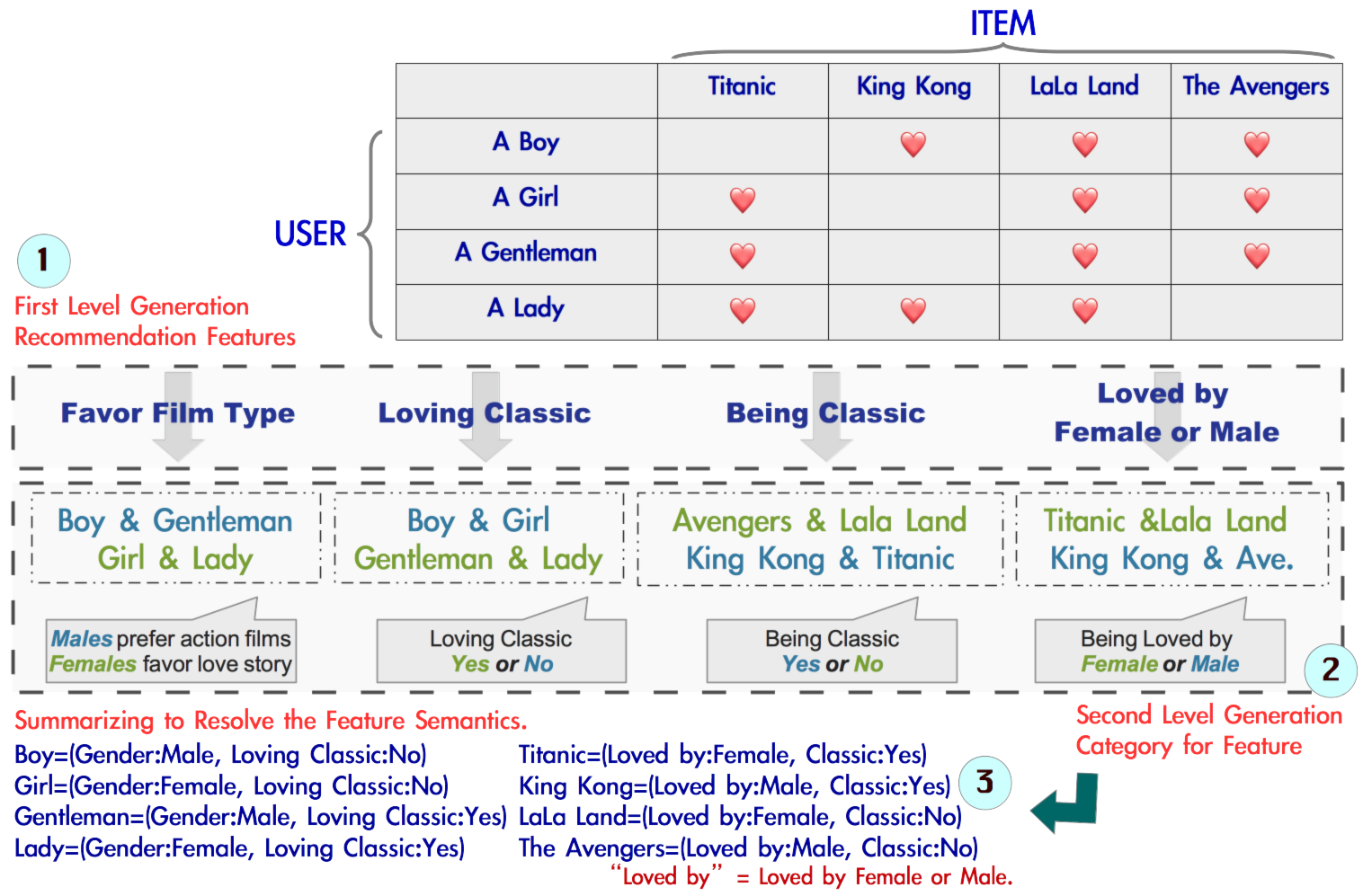}
	\caption{The illustration for the generative process of SAR. The users/items are semantically clustered from multiple views. Semantic features are generated from the first-level generative process, while the category in each feature, is generated from the second-level generative process. \textit{``Being Classic''}, \textit{``Gender'', etc}. are semantic features while \textit{Male}, \textit{Yes, etc}. are categories for the corresponding features.}
\end{figure*}

In this paper, we address a novel problem in recommendation: \textit{Will recommendation be improved by representing users and items with richer latent semantics?} Latent semantics refers to \textit{\textbf{the features or attributes for a user or an item, which are not directly observable}}. For instance, a user can be semantically described as \textit{(Age:Middle, Gender: Male, Occupation:Engineer, Action Film Fan:No...)}, while an item can also be represented in the same way, namely \textit{(Age Rage:Adult, Favored by Male:Yes, Welcomed by Engineer:Yes, Action Film:Yes, ...)}. Under this formulation, semantic matching between a user and an item would be more effective for recommendation than traditional matrix completion algorithms.

In comparison to purely real-valued vectorial representations, this semantic representation indicates the feature of a user/item in a latent semantic form such as LDA \cite{blei2003latent}. Then, with the semantic representations, recommendation can be more easily approached as a \textit{\textbf{semantic matching}} issue. In other words, if we can simultaneously represent user and item in the same semantic space, rating inference is simply to compute the similarity between the corresponding features of a user and those of an item. As shown in Figure \ref{CoreIdeaFigure}, the semantic features of the user \textit{``Gentleman''} are consistent with those of the movie \textit{``King Kong''}, which makes a ``to recommend'' rating.

Therefore, we propose a two-level hierarchical generative model \textbf{SAR} to discover and exploit the feature semantics merely from the rating matrix. At the first level of our model, we generate some features such as \textit{Action Film, Gender, etc}. At the second level of our model, we assign a corresponding category to each feature for users/items. Taking the example of movie \textit{``Quick and the Dead''}, we assign \textit{Yes} in the \textit{Action Film} feature, \textit{Youth} in the \textit{Loved by Youth or Elder} feature and so on. In this manner, users/items are semantically organized in a multi-view clustering form as shown in Figure \ref{CoreIdeaFigure}, which is a novel unsupervised paradigm. It is noting that, the semantics are learned in a latent form. Actually in our model, the feature and category are indicated as the vectors of cluster membership, rather than the linguistic lexicons (e.g. \textit{“Yes”, “Youth”}). Besides, the observable words of feature and category will be addressed in future work.

\textbf{Contributions.} This paper proposes a semantic analysis method for recommendation systems \textbf{(SAR)}, which applies a two-level hierarchical generative process that globally allocates semantic features and then locally assigns categories in each feature for users/items. Experimental results on real-world datasets show that our model consistently outperforms the state-of-the-art baselines.

\section{Related Work}
Existing recommendation methods can be roughly classified into four categories: \textit{matrix factorization, neighborhood based method, regression based method and social information integration  method}. Notably, the first three methods are all based on rating matrix completion.

\textbf{Matrix Factorization} is a conventional paradigm for recommendation. This methodology firstly applies a factorization form on the rating matrix $M \approx UV$ to get the factorization matrices $U$,$V$ and then multiplies $UV \approx \hat{M}$ to estimate the missing ratings, where $\hat{M}$ is the estimated rating matrix and $U$/$V$ is the user/item-related latent factor matrix respectively. Since this branch addresses different assumptions on $U$ and $V$, these methods fall into four primary subcategories according to the applied assumptions. \textit{(1.) Basic matrix factorization} generally emphasizes latent factors as being non-negative,  such as  NMF \cite{Wang2013Nonnegative}, SVP \cite{Meka2009Guaranteed}, MMMF \cite{Rennie2005Fast}, PMF \cite{Mnih2012Probabilistic}.  \textit{(2.) Combination of matrix factorization with neighborhood-based modeling}, such as CISMF \cite{Guo2015A}. \textit{(3.) Matrix factorization with complex rank assumptions} explores the rank features and generalization ability to enhance the factorization, such as LLORMA \cite{Lee2016LLORMA}\cite{Ko2015Multi}, R1MP \cite{Wang2014Rank}, SoftImpute \cite{Rahul2010Spectral}, \cite{Ganti2015Matrix}, \cite{Zhang2013Localized}, \cite{Kir2012The}. \textit{(4.) Matrix factorization with discrete assumptions} treats the output of the rating matrix as discrete values to avoid noise and obtain more interpretations, such as ODMC \cite{Huo2016Optimal}.

\textbf{Neighborhood Based Method} is one of the most classical approaches, assuming that the similar items/users hold similar rating preference. There are three main variants including item-based, user-based and global similarity, surveyed in \cite{Guo2015A} and \cite{Ricci2011Recommender}.

\textbf{Regression Based Method} is formulated as a regression problem, such as regression for graph GRALS \cite{Cai2011Graph},  blind regression \cite{Song2016Blind}, Riemannian manifold based regression \cite{Vandereycken2013Low}, and others \cite{Davenport20141}.

\textbf{Social Information Integration} leverages social information to enhance recommendation such as relationship between users, personalized profiles or movies' attributes. There list some latest researches: SR \cite{Ma2013An}, PRMF \cite{Liu2016Learning}, geo-specific personalization \cite{Liu2014Personalized}, social network based methods \cite{Deng2014Social} and other social context integration methods \cite{Wang2014Recommendation}.

\section{Methodology}
\subsection{Model Description}
A two-level hierarchical generative process is applied to discover and leverage the feature semantics as shown in Figure~\ref{GenProcess}. All the parameters of $\mathcal{P}(z|u, f)$, 
$\mathcal{P}(y|t,f)$, $\mathcal{P}(\hat{p}| z, y, f)$ are learned by the training procedure. Notably, $\mathcal{P}(f)$, $\mathcal{P}(u)$, $\mathcal{P}(t)$ are uniformly distributed, indicating that they can be safely omitted with simple mathematical manipulation.

\begin{figure}[H]
	\fbox{
		\textit{
			\parbox{0.965\linewidth}{
				For each user-item pair $(u, t) \in \mathcal{D}_{train}$:\\
				For each preference $p \in \{ 1, ..., |R| \}$:
				\begin{enumerate}
					\item \textbf{(First-Level)} \\ Draw a feature $f_n$ from $\mathcal{P}(f_n|u,t) $:
					\begin{enumerate}
						\item \textbf{(Second-Level)} Draw a user-specific category $z_n$ from 
						$ \mathcal{P}(z_n | u, f_n)$.
						\item \textbf{(Second-Level)} Draw an item-specific category $y_n$ from
						$ \mathcal{P}(y_n | t, f_n)$.
						\item \textbf{(Second-Level)} Draw a preference strength $\hat{p}$ from
						$ \mathcal{P}(\hat{p} | z_i, y_n, f_n, u, t)$.
					\end{enumerate}
					\item \textbf{(Rating Generation)} \\ Draw the rating $r$ from a soft-max distribution
					$$ \mathcal{P}(r | u, t) \propto e^{\hat{p} \omega_u \omega_t} $$
				\end{enumerate}
	}}}	
	\caption{The generative process of SAR. $\mathcal{D}_{train}$ is the training set. $|R|$ is the  maximum of ratings. }
	\label{GenProcess}
\end{figure}

Globally, we generate the rating $r$ for each user-item pair $(u, t)$ from a soft-max distribution (Gibbs distribution) where $\omega_u$,$\omega_t$ are user/item-relevant parameters. We denote the sense of each entry in the soft-max distribution as preference $p$, whose range is the same as that of all possible ratings, namely $\{1,...,|R|\}$.

Regarding the first-level, we generate a feature randomly from the uniform distribution $\mathcal{P}(f_n|u,t)$. Furthermore, we suppose each feature is equally important. For example, we can hardly distinguish which one is more important between \textit{Gender} and \textit{Occupation}.

Regarding the second-level, we generate a user/item-specific category and a preference strength for each feature. Obviously, the meaning of category depends on feature, since different categories are semantically differentiated under different features. For instance, the first category $(z = 1)$ under the feature of \textit{Gender} is \textit{Male}, while that under the feature of \textit{Occupation} is \textit{Student}. Intuitively, the preference strength $\hat{p}$ depends on both the user/item-specific category and the feature.

\begin{figure}
\centering
\includegraphics[width=0.8\linewidth]{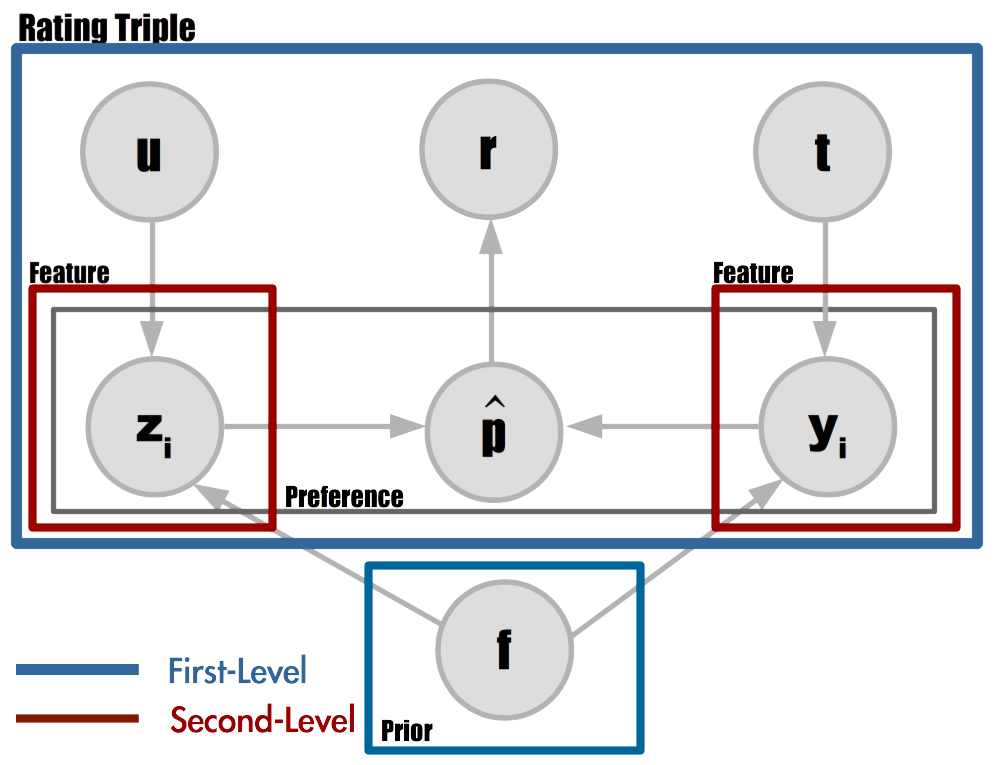}
\caption{The probabilistic graph of SAR. All the symbols and plates are explained in Figure \ref{GenProcess}.}
\label{PG}
\end{figure}

Besides, the features of user and item occurring in the same rating entry should be semantically proximal. Briefly speaking, the category distribution $\mathcal{P}(z_n|u,f_n)$ and $\mathcal{P}(y_n|t,f_n)$ are supposed to be consistent as required by the demand of semantic matching. Taking Figure \ref{CoreIdeaFigure} as an example, the \textit{``Lady''} favors \textit{``Titanic''}, which indicates she loves the	classic movies while the \textit{``Titanic''} is a classic one. Therefore, a Laplace prior is applied in $\mathcal{P}(\hat{p}|z_n, y_n, f_n)$ to formulate this observation as below:
\begin{eqnarray}
\mathcal{P}(\hat{p}|z_n, y_n, f_n, u, t) =
\tau_{\hat{p}|z_n, y_n, f_n} e^{-\frac{| \mathcal{P}(z_n | u, f_n) - \mathcal{P}(y_n | t, f_n)|}{\sigma}} \label{e0}
\end{eqnarray}
where $\tau_{\hat{p}|z_n, y_n, f_n}$ is tabular model parameters which can be
tuned in the learning process and $\sigma$ is the hyper-parameter.

\begin{figure*}
	\begin{eqnarray}
	&& \mathcal{P}(\hat{p}, z_n, y_n, f_n|u,t)  = \mathcal{P}(z_n|u, f_n) \mathcal{P}(y_n|t, f_n) \mathcal{P}(\hat{p}|z_n, y_n, f_n, u, t) \label{e1}
	\end{eqnarray}
	\begin{small}
		\fontsize{8pt}{0}
		\begin{eqnarray}
		& \mathcal{P}(\hat{p}|u,t) & =
		\overbrace{\sum_{n=1}^{|F|} \underline{\mathcal{P}(f_n|u,t)}
			\overbrace{\sum_{i,j=1}^{|C|}  \underline{\mathcal{P}(z_n=i|u, f_n) \mathcal{P}(y_n=j|t, f_n)} \mathcal{P}(\hat{p}|z_n=i, y_n=j, f_n, u, t) 
			}^{Second-Level:~Category~Mixture}
		}^{First-Level:~Feature~Mixture} \label{e2} \\
		& & =
		\sum_{n=1}^{|F|} \mathcal{P}(f_n|u,t)
		\sum_{i,j=1}^{|C|}  \mathcal{P}(z_n=i | u, f_n) \mathcal{P}(y_n=j|t, f_n) \tau_{\hat{p}|z_n, y_n, f_n} e^{-\frac{| \mathcal{P}(z_n =i| u, f_n) - \mathcal{P}(y_n=j | t, f_n)|}{\sigma}} \nonumber \\ \label{e3}
		\end{eqnarray}
	\end{small}	
\end{figure*}

Figure \ref{PG} presents the probabilistic graph of SAR, with which we can compute the joint probability as (\ref{e1}), (\ref{e2}) and (\ref{e3}), where $|F|$ and $|C|$ are the number of features and categories, respectively. It is noteworthy that, $\mathcal{P}(f_i|u,t)$ are uniformly
distributed.

Figure \ref{PG} presents two inference tasks of SAR: rating prediction and user/item-specific category prediction. Regarding the first task, rating is estimated as the expectation of soft-max distribution, as formulated in (\ref{e4}).
\begin{eqnarray}
r_{|u,t} \doteq \mathbb{E}_{r|u,t}(r) =  \frac{\sum_{p=1}^{|R|} p \times e^{\hat{p} \omega_u \omega_t}}{\sum_{p=1}^{|R|} e^{\hat{p} \omega_u \omega_t}} \label{e4}
\end{eqnarray}
where $\hat{p}$ is the corresponding preference strength. Regarding the second task, category distribution is calculated respectively as (\ref{e5}) and (\ref{e6}), which are the direct results of (\ref{e1}) by applying the sum rule. Notice that all the $\mathcal{P}(f_i)$, $\mathcal{P}(u)$ and $\mathcal{P}(t)$ are uniform distributions.
\begin{eqnarray}
\mathcal{P}(z_n|u) = \sum_{t, f_n, y_n, \hat{p}} \mathcal{P}(\hat{p}, z_n, y_n, f_n|u,t)\label{e5} \\
\mathcal{P}(y_n|t) = \sum_{u, f_n, z_n, \hat{p}} \mathcal{P}(\hat{p}, z_n, y_n, f_n|u,t) \label{e6} 
\end{eqnarray}

\subsection{Objective \& Training}
We formulate the objective function as the mean square error between the predicted and golden ratings. Furthermore, we also apply the regularization term in the objective function.
\begin{eqnarray}
\mathcal{L} & & = \sum_{(u,r,t) \in \mathcal{D}_{train}} (r_{|u,t}  - r)^2 +  \\
& & \lambda \left\{ \sum_{n=1}^{|F|}||\mathcal{P}(z_n|u, f_n)||_2^2 +  \sum_{n=1}^{|F|}||\mathcal{P}(y_n|t, f_n)||_2^2 \nonumber \right\} 
\end{eqnarray}
where $D_{train}$ is the training dataset and $r_{|u,t}$ is the predicted rating using (\ref{e3}). $||~||_2^2$ is the $l_2$ norm.

The parameters are as follows: $\mathcal{P}(z_n | u, f_n)$, $\mathcal{P}(y_n|t, f_n)$, $\tau_{\hat{p}|z_n, y_n, f_n}$, $\omega_u$, $\omega_t$, all of which are learned by minimizing the objective function $\mathcal{L}$. For a more efficient and facilitating solution, a moment-based gradient method AdaDelta \cite{Zeiler2012ADADELTA} is adopted with hyper-parameters: moment factor $\eta$ and RMSE factor $\epsilon$.

Regarding the efficiency, theoretically, the time complexity of SAR is $O(|F||C|^2)$, while practically the running time is present in Table \ref{time-table}, which illustrates our method is indeed efficient.

\subsection{Analysis from the Clustering Perspective}
Referring to the Equation (\ref{e2}), SAR essentially adopts a hierarchical mixture form, where the underline corresponds to the first- and second-level mixture factors, respectively. Because mixture leads to grouping, SAR can be further analyzed from the perspective of clustering, as a multi-view methodology. In the second-level, or in each clustering view, the generative process groups the users/items, according to the feature semantics of this view. These semantic features originate from the first-level process, mathematically corresponding to all the probabilistic terms involved with $f_n$. On the other hand, the first-level process adaptively adjusts different feature semantics according to the feed-back information from the second-level, precisely speaking as $\mathcal{P}(z_{1...n}, y_{1...n}, f|u,r,t)$. \textit{In conclusion, users/items are semantically differentiated in a multi-view clustering form. Thus, with the exploitation of multi-view nature, recommendation can be promoted by semantic matching.}

In Figure \ref{CoreIdeaFigure}, there is a user-item rating matrix. Traditional clustering process (e.g. K-MEANS) for users/items is ambiguous, because there is always various clustering aspects, such as grouping by \textit{Gender}, or by \textit{Loving Classic, etc}. However, in our model, different semantic features \textit{(e.g. Gender, Loving Classic)} are generated from the first-level process. Therefore, to approach the clustering more easily, the second-level addresses the grouping of users/items, according to one exact feature. Specifically in the example of Figure \ref{CoreIdeaFigure}, within the view/feature of \textit{Gender}, the user \textit{Boy} belongs to \textit{Male} cluster, while within that of \textit{Loving Classic}, this user belongs to the \textit{No} cluster. Finally, summarizing each semantic feature, SAR discovers the feature semantics of the user \textit{Boy}, namely \textit{Boy = (Gender:Male, Loving
Classic:No)}. Notably, the feature semantics is a “latent” semantic information to empower recommendation systems.

\subsection{Analysis from the Semantic Perspective}
Once SAR discovered the feature semantics as shown in Figure \ref{CoreIdeaFigure}, the recommendation can be strengthened in the manner of semantic matching. It means it is sufficient to compare the semantic features between users and items to infer the ratings. For example, when predicting whether the \textit{Boy} would like \textit{``Titanic''}, we just compare the corresponding features of the user with those of the item. Then no match is found, so it is reasonable to predict a ``not to recommend'' rating. Conversely, when the girl rates \textit{``LaLa Land''}, semantic matching would reasonably derive a ``to recommend'' rating. Mathematically, semantic matching is modeled by the Laplace prior of Equation (\ref{e0}).

\section{Experiment}
\subsection{Datasets.}
We conduct our experiments on two public benchmark datasets: MovieLens 100K (\textit{ML100K} for short) and MovieLens 1M (\textit{ML1M}). ML100K consists of 100,000 ratings given by 943 users to 1,682 movies, while 93.69\% of entries in the rating matrix is empty, while ML1M contains 1,000,209 ratings given by 6,040 users to 3,706 movies, and 95.53\% of entries are missing.

\subsection{Performance Evaluation}
The rating prediction is a traditional benchmark task, which concerns the predictive ability for matrix completion. This task directly benefits many recommendation applications, such as item-based recommendation \cite{Zhang2013Localized}.

\begin{table}[t]
	\caption{Rating prediction results on ML100K (top) and ML1M (bottom), measured by MAE and RMSE.}
	\label{result-table-100K}
	\centering
	\renewcommand\arraystretch{1.05}
	\begin{tabular}[t]{c<{\centering}|p{1.82cm}<{\centering}| p{1.82cm}<{\centering} | p{1.82cm}<{\centering}}
		\hline \textbf{Sparsity} & \textbf{Method} & \textbf{MAE} & \textbf{RMSE} \\
		\hline
		\hline
		\multirow{10}{*}{$\rho=80\%$} & PMF & 0.7522 & 0.9667 \\
		\cline{2-4} & NMF & 0.7724 & 0.9874 \\
		\cline{2-4} & MMMF & - & 0.9800 \\
		\cline{2-4} & User-Based & 0.7370 & 0.9440 \\
		\cline{2-4} & CISMF & 0.7279 & 0.9268 \\
		\cline{2-4} & ODMC & 0.7033 & 0.9609 \\
		\cline{2-4} & GRALS & - & 0.9450 \\
		\cline{2-4} & SR & 0.7298 & 0.9218 \\
		\cline{2-4} & PRMF & 0.7215 & 0.9135 \\
		\cline{2-4} & \textbf{\textit{SAR(Ours)}} & \textit{\textbf{0.6772}} (\textit{\textbf{0.7133}})$^*$ & \textit{\textbf{0.9069}} \\
		\hline
		\hline
		\multirow{5}{*}{$\rho=50\%$} & SVP & - & 0.968 3 \\
		\cline{2-4} & R1MP & - &  1.0168 \\
		\cline{2-4} & SoftImpute & - & 1.0354 \\
		\cline{2-4} & \textbf{\textit{SAR(Ours)}} & \textit{\textbf{0.6954}}
		\textit{\textbf{(0.7332)}}$^*$ & \textit{\textbf{0.9280}} \\
		\hline
	\end{tabular}
	\label{result-table-1m}
	\centering
	\vspace{2px}
	\renewcommand\arraystretch{1.05}
	\begin{tabular}[t]{c<{\centering}|p{1.82cm}<{\centering}| p{1.82cm}<{\centering} | p{1.82cm}<{\centering}}
		\hline \textbf{Sparsity} & \textbf{Method} & \textbf{MAE} & \textbf{RMSE} \\
		\hline
		\hline
		\multirow{7}{*}{$\rho=80\%$} & PMF & 0.7306 & 0.9234 \\
		\cline{2-4} & NMF & 0.7286 & 0.9203 \\
		\cline{2-4} & MMMF & - & 0.8810 \\
		\cline{2-4} & User-Based & 0.7030 & 0.9050 \\
		\cline{2-4} & LLORMA & 0.6941 & 0.8911 \\
		\cline{2-4} & ODMC & 0.6583 & 0.9371 \\
		\cline{2-4} & \textbf{\textit{SAR(Ours)}} & \textit{\textbf{0.6436}}
		\textit{\textbf{(0.6886)}}$^*$ & \textit{\textbf{0.8715}} \\
		\hline
		\hline
		\multirow{5}{*}{$\rho=50\%$} & SVP & - & 0.9085 \\
		\cline{2-4} & R1MP & - & 0.9595 \\
		\cline{2-4} & SoftImpute & - & 0.8989 \\
		\cline{2-4} & Blind Regression & - & 0.9220 \\
		\cline{2-4} & \textbf{\textit{SAR(Ours)}} & \textbf{\textit{\textbf{0.6543}}}
		\textit{\textbf{(0.6961)}}$^*$ & \textit{\textbf{0.8827}} \\
		\hline
	\end{tabular}
\end{table}

\textbf{Protocol. \& Metric.}
We evaluate how SAR works on the datasets of different sizes and sparsity. Hence, we vary the ratio of \textit{training to test data $\rho$} to evaluate our algorithm and compare with baselines. For example if $\rho = 80\%$, 80\% of the observed ratings are randomly sampled for training, and the remaining 20\% observed ratings are used for test. Two widely used measures are employed for prediction evaluation, namely \textit{Root Mean Square Error (RMSE)} and \textit{Mean Absolute Error (MAE)} \cite{Lee2016LLORMA}.

\let\thefootnote\relax\footnotetext{$^*$For a better MAE score, we apply a trick that to round the prediction. Note that, at least, ODMC has applied the similar trick. But, to be strict, we have reported the non-trick performance. }

\textbf{Baselines \& Implementation.}
We compare SAR with 14 state-of-the-art baselines, almost including all the major approaches introduced in Related Work. Specially, SR is a method that encodes the social information into matrix completion, and NMF constrained all the factored matrix positive. \textit{For a fair comparison with previously proposed methods, we directly reprint the results under the same setting from the literature.} On both datasets, the optimal configuration of SAR are as follows: feature number $|F|=10$, category number $|C|=10$, Laplace prior hyper-parameter $\sigma = 1.0$, regularization factor $\lambda=0.05$, moment factor $\eta=0.6$ and $\epsilon=1 \times 10^{-6}$. We train the model until convergence, but at most 400 rounds.

\begin{figure*}
	\centering
	\includegraphics[width=\linewidth]{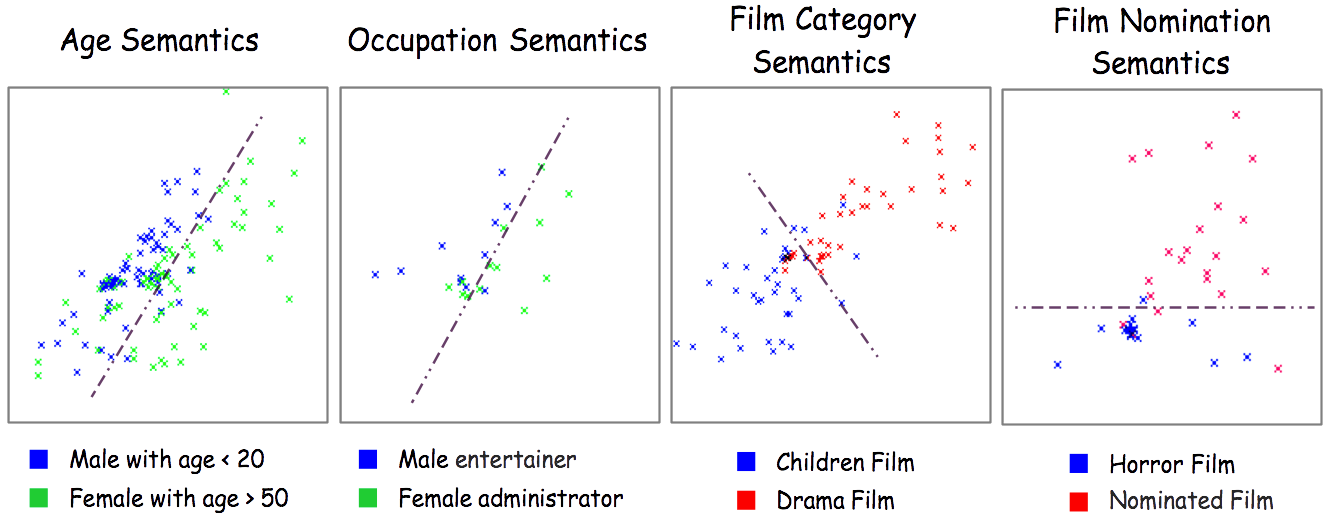}
	\caption{The demonstration of semantic evaluation. Each point indicates a distribution (namely $\mathcal{P}(z_n|u)$ or $\mathcal{P}(y_n|t)$), which is plotted after projected by PCA. The line is the boundary between two different categories. The top text indicates the feature and the bottom legend represents the categories. The left two figures are user-specific and the right two are item-specific. }
	\label{SemanticEvaluationScatter}
\end{figure*}

\textbf{Prediction Accuracy.} Evaluation results are shown in Table \ref{result-table-1m}. We can make the following statements. 
\begin{enumerate}
	\item SAR outperforms all the baselines, justifying the effectiveness of our model. 
	\item Theoretically, the effectiveness stems from the semantics-specific modeling of SAR,
	justifying that SAR can discover and utilize more semantics than social information integration methods do.
\end{enumerate}

\begin{table}[t]
	\caption{Average computational time for one round (measured in seconds). The experiments are conducted on i7-6900K (3.87 GHz) with 32GB Memory.}
	\centering
	\label{time-table}
	\renewcommand\arraystretch{1.1}
	\begin{tabular}{c|c|c|c|c}
		\hline
		\textbf{Dataset} & \textbf{NMF} & \textbf{PMF} & \textbf{MMMF} & \textbf{SAR(Ours)} \\
		\hline
		\hline
		\textit{ML100K} & 0.9 & 4.4 & 133.2 & 5.5 \\
		\textit{ML1M} & 5.5 & 16.9 & 405.5 & 38.1 \\
		\hline
	\end{tabular}
\end{table}

\textbf {Time Complexity.} Computation efficiency is evaluated as shown in Table \ref{time-table}, from which we can observe that SAR is as efficient as NMF.  As to the trade-off between efficiency and effectiveness, we propose the metric of RMSE improvement over NMF per second: $-\frac{RMSE(algorithm) - RMSE(NMF)}{Time(algorithm) - Time(NMF)}$ \textit{(the bigger, the better)}. In terms of this measurement, the result of PMF is $6.00 \times 10^{-3}$ and that of SAR is $1.60 \times 10^{-2}$ on ML100K, while $2.74 \times 10^{-4}$, $1.50 \times 10^{-3}$ on ML1M for PMF and SAR, respectively. The comparison illustrates that SAR is a better time-performance trade-off method, which is more practical for real-world applications.

\begin{figure}[t]
	\centering
	\includegraphics[width=0.6\linewidth]{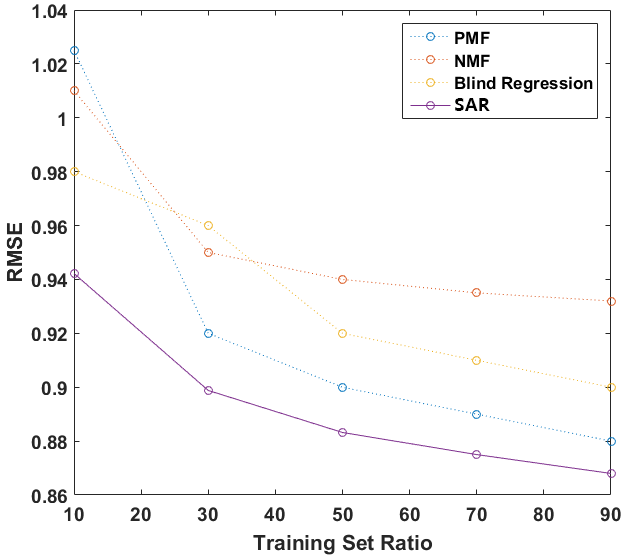}
	\caption{RMSE of PMF, NMF, Blind Regression on the ML1M dataset when varying the training set ratio $\rho$.}
	\label{sparisty-figure}
\end{figure}

\textbf{Sparsity.} We change the $\rho$ to verify the effect of sparsity and the results are shown in Figure \ref{sparisty-figure}. We find that all the methods improve as the training set size increases while SAR consistently outperforms the other baselines. For the case of blind regression, where $\rho$ varies from 90\% to 30\%, the RMSE degrades 0.050, while for SAR, the result is 0.032. This comparison demonstrates that SAR is less sensitive to data sparsity than blind regression which is a strong baseline in this aspect.

\begin{figure}[t]
	\centering
	\includegraphics[width=\linewidth]{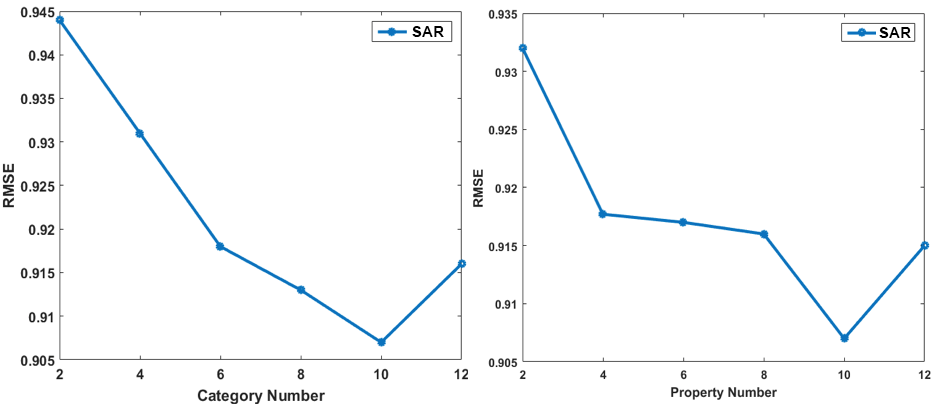}
	\caption{The effect of hyper-parameters. The y-axis is the RMSE on ML100K with  $\rho=80\%$. The x-axis means varying the corresponding hyper-parameters with the optimal configuration of other settings as introduced in \textbf{Baselines \& Implementation.}}
	\label{hyper}
\end{figure}

\textbf{Hyper-Parameters.} As illustrated in Figure \ref{hyper}, under the normal settings,  SAR is insensitive to hyper-parameters. Concretely, when the category number $|C|$ varies between 6 and 12, the performance (RMSE) changes within a magnitude of 0.015, and when the feature number $|F|$ changes between 6 and 12, the performance  (RMSE) also varies within a small magnitude of 0.010. This comparison justifies the robustness of SAR. 

\subsection{Semantic Evaluation}
In this subsection, we conduct an experiment to testify our statement about feature semantics. GroupLens is chosen since it provides the user/item profiles (e.g. gender, occupation, film type) for ML100K, which can facilitate the semantic analysis of our model.

\textbf{Experimental Setting.} The empirical results are shown in Figure \ref{SemanticEvaluationScatter}, which is proceeded with the following steps. First, we calculate each distribution according to $(\ref{e5})$ and $(\ref{e6})$ for each user/item and then represent a user/item with the corresponding distribution (namely $\mathcal{P}(z_i|u)$ or $\mathcal{P}(y_j|t)$) as a point in high-dimensional space $\mathbb{R}^{|C|}$. Second, PCA is conducted to project $|C|$-dimensional points to a 2D plane. At last, the corresponding attributes from the profile data are manually analyzed to color the points. For clarity, a line is painted artificially to illustrate the boundary.

\textbf{Results.}  We can easily see that SAR discovers and leverages the feature semantics appropriately, as shown in Figure \ref{SemanticEvaluationScatter}.  

Concretely, regarding the left figures, user-specific semantics are identified as a combination of gender and age attributes. In fact, only one single user attribute can hardly distinguish a preference (i.e., a rating), but a combination of user attributes is capable to semantically identify the preference. For example in the most left sub-figure, the blue points indicate the teenager boys and the green points denote the housewives or senior business women. 
Since the features in terms of recommendation are different, the two point groups are discriminated to a large extent. 

Additionally, we found the item-relevant semantics are visualized as a transition between attributes components in SAR.  In the sub-figure of \textit{Children - Drama Film}, the proximity of the boundary lays some \textit{Children Drama} films, such as \textit{``Little Prince (1995)''} and \textit{``Secret Garden (1993)''}. But conversely in the rightest sub-figure, there are almost no films in the transition area, because in the ML100K dataset, there is not a nominated horror film at all. This phenomenon illustrates the semantic effectiveness of SAR, which can benefit the recommendation by semantic matching.

\section{Conclusion}
In order to discover and utilize latent feature semantics, we propose a semantic analysis approach for recommendation, \textbf{(SAR)}. The model applies a two-level hierarchical generative process that assigns semantic features and categories for user and item. Experimental results on benchmark datasets demonstrate the effectiveness of our proposed methods .

\bibliography{SAR}
\bibliographystyle{named}	
\end{document}